\documentclass[sigconf]{acmart}

\usepackage{listings}
\usepackage{xcolor}
\usepackage{tikz}
\usepackage{makecell}
\usetikzlibrary{arrows.meta, positioning}

\lstdefinelanguage{MLIR}{
  sensitive=true,
  morecomment=[l]{//},
  morecomment=[s]{/*}{*/},
  morestring=[b]",
  emph={builtin, func, device, hls, scf, arith}, 
  emphstyle=\color{gray},
  morekeywords={
    module, return,
    alloc, data_acquire, data_release,
    kernel_create, kernel_launch, kernel_wait,
    interface, axi_protocol, constant, index_cast,
    load, store, addf, subi, for 
  },
}

\lstdefinestyle{mlir_listing}{
  language=MLIR,
  basicstyle=\ttfamily\scriptsize,
  keywordstyle=\color{blue!60!cyan}\bfseries,
  emphstyle=\color{gray}\bfseries,
  commentstyle=\color{gray},
  stringstyle=\color{green!40!black},
  breaklines=true,
  showstringspaces=false,
  columns=flexible,
  numbers=none,
  frame=none,
  xleftmargin=0pt,
  framexleftmargin=0pt,
  escapeinside={(*@}{@*)} 
}

\newcommand{\mlirop}[2]{\textbf{\textcolor{gray}{#1}\textcolor{blue!60!cyan}{.#2}}}

\NewDocumentCommand{\mlirinline}{ m m }{%
  \textbf{\textcolor{gray}{#1}\textcolor{blue!60!cyan}{.#2}}%
}

\AtBeginDocument{%
  }  

\makeatother

\copyrightyear{2025}
\acmYear{2025}
\setcopyright{cc}
\setcctype{by}
\acmConference[SC Workshops '25]{Workshops of the International Conference for High Performance Computing, Networking, Storage and Analysis}{November 16--21, 2025}{St Louis, MO, USA}
\acmBooktitle{Workshops of the International Conference for High Performance Computing, Networking, Storage and Analysis (SC Workshops '25), November 16--21, 2025, St Louis, MO, USA}
\acmDOI{10.1145/3731599.3767485}
\acmISBN{979-8-4007-1871-7/2025/11}

\begin{document}

\title{An MLIR pipeline for offloading Fortran to FPGAs via OpenMP}

\author{Gabriel Rodriguez-Canal}
\email{gabriel.rodcanal@ed.ac.uk}
\orcid{0009-0005-0511-3922}
\affiliation{%
  \institution{EPCC, The University of Edinburgh}
  \city{Edinburgh}
  \country{United Kingdom}
}

\author{David Katz}
\email{david.katz@ed.ac.uk}
\orcid{0009-0003-7387-6169}
\affiliation{%
  \institution{EPCC, The University of Edinburgh}
  \city{Edinburgh}
  \country{United Kingdom}
}

\author{Nick Brown}
\email{n.brown@epcc.ed.ac.uk}
\orcid{0009-0005-0511-3922}
\affiliation{%
  \institution{EPCC, The University of Edinburgh}
  \city{Edinburgh}
  \country{United Kingdom}
}

\renewcommand{\shortauthors}{Rodriguez-Canal et al.}

\begin{abstract}
With the slowing of Moore’s Law, heterogeneous computing platforms such as Field-Programmable Gate Arrays (FPGAs) have gained increasing interest for accelerating HPC workloads. In this work we present, to the best of our knowledge, the first implementation of selective code offloading to FPGAs via the OpenMP target directive within MLIR. Our approach combines the MLIR OpenMP dialect with a High-Level Synthesis (HLS) dialect to provide a portable compilation flow targeting FPGAs. Unlike prior OpenMP FPGA efforts that rely on custom compilers, by contrast we integrate with MLIR and so support any MLIR-compatible front end, demonstrated here with Flang. Building upon a range of existing MLIR building blocks significantly reduces the effort required and demonstrates the composability benefits of the MLIR ecosystem. Our approach supports manual optimisation of offloaded kernels through standard OpenMP directives, and this work establishes a flexible and extensible path for directive-based FPGA acceleration integrated within the MLIR ecosystem.
\newline Code available in \url{https://github.com/xdslproject/ftn/tree/main}.
\end{abstract}

\ccsdesc[500]{Hardware~Reconfigurable logic and FPGAs}
\ccsdesc[500]{Software and its engineering~Compilers}


\keywords{MLIR, LLVM, OpenMP, FPGAs, AMD U280, High Level Synthesis, HPC}

\maketitle

\section{Introduction}
Driven by the ever growing demand to model more complex systems and reduced time to solution, heterogeneous computing platforms such as Field-Programmable Gate Arrays (FPGAs) have emerged as potential accelerators for scientific and high-performance computing (HPC) workloads. FPGAs are especially relevant given their low power draw and the increased focus in HPC on sustainable supercomputing. FPGAs offer fine-grained customisation and power efficiency, making them particularly suited for codes with irregular memory access patterns or those not solely bound by raw compute performance \cite{brown2020exploring}. However, programming FPGAs remains a major challenge, especially for HPC developers accustomed to directive-based models like OpenMP.

FPGA vendors such as AMD, formerly Xilinx, provide high-level synthesis (HLS) tools that simplify Register Transfer Level (RTL) generation but provide limited support for integration between the host and device within standard parallel programming models. A major challenge with HLS is that programmers must learn new annotations and a new programming model before they are able to port their codes. By contrast, OpenMP \cite{chandra2001parallel} is a standard parallel programming technology that the majority of scientific programmers are already familiar with. Since target offload was introduced in version 4.0, OpenMP has supported the acceleration of loops on devices such as GPUs. There have been numerous efforts to leverage OpenMP to offload codes to FPGAs \cite{mayer2019openmp} but these efforts such as OmpSS@FPGA \cite{de2021ompss}, Nymble \cite{huthmann2020openmp} and ORKA-HPC \cite{mayer2021orka}, require significant compiler development and some also involve non-standard OpenMP extensions. This ultimately limits portability, imposes tool chain maintenance challenges and results in user risk of adoption. Furthermore, these existing OpenMP FPGA flows are all focused on C/C++ which, given the prevalence of Fortran in HPC, is a major limitation as Fortran codes must first be ported to a new language.

In this paper we present, to the best of our knowledge, the first integration of selective OpenMP target offloading to FPGAs within the Multi-Level Intermediate Representation (MLIR) framework. Built around the core upstream dialects and driven by the Flang Fortran compiler, we demonstrate a flow that generates LLVM-IR that is provided to AMD's HLS backend tooling. The fundamental hypothesis that we test in this paper is that the composability provided by MLIR is uniquely advantageous as it enables one to significantly reduce the development effort in supporting OpenMP to FPGA offloading. This is due to being able to leverage a range of existing building blocks that we highlight and describe.

This paper is structured as follows; in Section \ref{sec:bg} we explore the background to this work by describing FPGAs in more detail, introducing the LLVM, MLIR and Flang compiler technologies and surveying related work in programming FPGAs for HPC. Section \ref{sec:our_flow} then introduces our MLIR-based compiler flow for driving the offload of Fortran loops to FPGAs via OpenMP, before evaluating this in Section \ref{sec:eval}. We then draw conclusions and describe further work in Section \ref{sec:conclusions}.

\section{Background and related work}
\label{sec:bg}

\begin{figure*}[htb]
\centering
 \includegraphics[width=\textwidth]{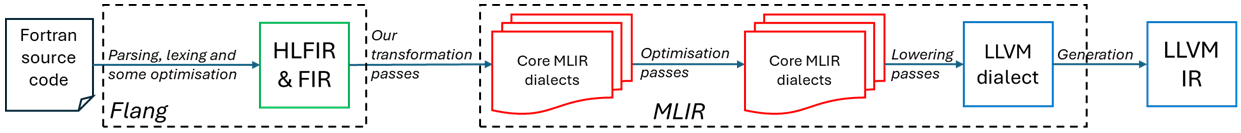}
\caption{Illustration of flow developed in \cite{brown2024fully} to lower Flang to core dialects and then generate LLVM-IR from \cite{brown2024fully}.}	
\label{fig:flang-flow}
\end{figure*}

Field Programmable Gate Arrays (FPGAs) are a reconfigurable architecture that provide a large number of configurable logic blocks sitting within a sea of configurable interconnect. Other on-chip blocks, such as block RAM (BRAM) and UltraRAM (URAM) memory, and Digital Signal Processing (DSP) slices for arithmetic provide great versatility and potential for performance. Furthermore, the low clock rate, typically between 200 Mhz and 400 Mhz, results in low power draw and potential for high energy efficiency \cite{brown2020exploring}. Given these benefits it might seem surprising that the adoption of this technology in scientific computing has, until now, been limited. Historically this was due to the immaturity of both the hardware and software ecosystems, however in recent years vendors have made significant investments. This has resulted in FPGAs which are much more powerful and hence a more realistic proposition for HPC \cite{nguyen2022fpga}, along with improvements to the software ecosystem.

Vitis \cite{xilinx2022vitis}, developed by Xilinx which is now part of AMD, has been a major success in the software ecosystem. Shifting the programming of FPGAs to become much more a question of software development, rather than hardware design, programmers write their code in C decorated with specialist pragmas. The High Level Synthesis (HLS) tooling then converts this into its Register Transfer Level (RTL) counterpart. Whilst HLS tools existed before Vitis, the benefit of this framework is the overarching integration as it packages everything together. Consequently, using the \emph{v++} command line tool one is able to compile their C code using HLS, integrate the resulting RTL into the shell running on the FPGA, and to drive the remaining synthesis, placement and routing steps to ultimately generate a bitstream that will program the device.

However, all these successes in making FPGA programming simpler only guarantee to provide correctness. In order to obtain best performance on FPGAs programmers must still make significant algorithmic changes in order to convert these to a dataflow form \cite{de2020transformations}. Furthermore, they must learn new tooling and techniques along with integration into their existing code. This is often non-trivial as programmers must write code running on the host to manage data transfers, kernel launching on the device and synchronisation of FPGA kernels and host code.

\subsection{LLVM, MLIR and Flang}

LLVM \cite{lattner2004llvm} is a collection of modular and reusable compiler and toolchain technologies designed to support the development of compilers for a wide range of programming languages and hardware architectures. It provides numerous language-specific frontends and architecture-specific backends, all connected through an intermediate representation known as LLVM-IR. For example, a frontend such as Clang, which handles C and C++, generates LLVM-IR that is then subsequently processed by a backend. These backends span a variety of targets, including CPUs, GPUs, and FPGAs. However, because LLVM-IR is a relatively low-level representation, each frontend must perform substantial work to translate source code into this form. This leads to duplicated effort across frontend implementations.

To address this challenge, MLIR (Multi-Level Intermediate Representation) was initially developed by Google and later released as open-source software. MLIR introduces a system of intermediate representation dialects along with transformations between them. Instead of targeting LLVM-IR directly, language frontends generate higher-level MLIR dialects, relying on shared infrastructure within MLIR to incrementally lower these representations toward LLVM-IR. Like LLVM-IR, MLIR is based on Static Single Assignment (SSA) form and a key advantage of MLIR is its support for combining multiple dialects within a single IR. This enables progressive lowering and separate manipulation of distinct abstraction levels, and such modularity facilitates greater reuse of compiler infrastructure across different frontends. Ultimately, MLIR has the potential to significantly reduce the effort required to build and maintain compilers and furthermore provides an extensible framework where developers can define their own custom dialects and transformations.

MLIR has gained substantial adoption in the past few years and is supported by an active and growing community. The MLIR ecosystem includes a wide range of core dialects, such as \emph{arith} for arithmetic operations and \emph{scf} for structured control flow including serial and parallel loops. The \emph{affine} dialect represents affine loop nests, \emph{memref} is provided for memory management and data access, \emph{func} for functions and subroutine calls and \emph{openmp} characterises OpenMP-based parallelism. 

One of the disadvantages of MLIR is the steep learning curve, where developers must leverage C++, understand LLVM concepts, work with the Tablegen format to describe dialects and keep track of the fast moving MLIR repository. This adds to the overhead involved in development. To this end xDSL \cite{fehr2025xdsl} was developed as a Python based compiler toolkit which is 1-1 compatible with MLIR. Providing the majority of standard MLIR dialects, as well as numerous experimental ones too, a major purposes of this toolkit is to enable rapid exploration and prototyping of MLIR
concepts. The MPI dialect is one such example of this, where it was first developed using xDSL in \cite{bisbas2024shared} being standardised in MLIR. We have used xDSL to develop the transformations and optimisations described in this paper, and as it is 1-1 compatible with MLIR it is trivial at any point to go into the MLIR ecosystem via \emph{mlir-opt}.

Flang \cite{flang} is the LLVM project's official Fortran frontend, the result of a complete redesign and rewrite of the original classic Flang compiler. Built from the ground up using MLIR, this new Flang aims to provide comprehensive support for the Fortran language standard, with the flexibility to accommodate future revisions of the language. As an integrated component of LLVM, Flang is under active development and evolving rapidly. After parsing and lexing, Flang generates IR based upon the HLFIR (High Level Fortran Intermediate Representation) and FIR (Fortran Intermediate Representation) dialects, before undertaking a series of transformation and optimisation passes on these dialects ultimately resulting in LLVM-IR. Consequently, Flang sits outside of the core MLIR ecosystem as it integrates with only a subset of the core dialects and HLFIR \& FIR are not part of MLIR itself, instead with Flang providing its own path to LLVM-IR.

In \cite{brown2024fully} the authors developed a lowering from HLFIR \& FIR into core MLIR dialects, such as \emph{memref} for handling variables and \emph{scf} for handling control flow. This flow is illustrated in Figure \ref{fig:flang-flow} and it was found that integrating with the rest of the MLIR ecosystem provides some performance benefits, as the wide range of MLIR transformations can be leveraged. Furthermore, there were also flexibility benefits resulting from gaining access to a wider range of other dialects and transformations that are part of MLIR.

\subsection{FPGA programming approaches}

AMD's HLS tooling is based upon Clang and they have open sourced their fork of the Clang C/C++ frontend, along with documenting the API of their backend. Considering the popularity of Fortran in HPC, where around 80\% of codes running on the Cray-EX ARCHER2 supercomputer are written in Fortran \cite{rodriguez2023fortran}, it is a considerable undertaking first rewriting Fortran code in C/C++ before being able to bring it to the FPGA. Consequently, \cite{rodriguez2023fortran} coupled LLVM-IR generated by Flang with the HLS backend, enabling Fortran HLS programming. This work enabled HLS pragmas to be specified in Fortran, where a preprocessor picks them out, translates them into function calls and a later stage in the pipeline then converts these into AMD's bespoke HLS LLVM-IR primitives. Furthermore, the flow had to provide compatibility between the latest LLVM-IR generated by Flang and AMD's backend which is version 7 of LLVM. 

However, whilst \cite{rodriguez2023fortran} solves the issue of porting code between Fortran and C/C++, programmers must still use HLS pragmas and manually couple their existing code with the device, having to explicitly write host code that handles data transfers and manage kernel execution on the FPGA. OmpSs@FPGA \cite{de2021ompss} developed FPGA offloading primarily with programmers leveraging the \emph{task} and \emph{target} directives of OpenMP. Supporting C and C++, programmers define tasks that are then scheduled and run on the FPGA. Whilst OmpSs@FPGA uses LLVM, it required a significant amount of bespoke compiler development and its own OmpSs runtime system, ultimately generating C++ code that is annotated with HLS pragrams and used as input to AMD's HLS tooling. 

Nymble \cite{huthmann2020openmp} is also built upon the LLVM and provides a compiler flow that maps OpenMP target regions to FPGAs. LLVM-IR is generated and used for HLS optimizations, and this is integrated with the Nymble HLS compiler to generate the corresponding RTL. Supporting OpenMP pragmas including \emph{target}, \emph{teams}, and \emph{parallel} this compiler uses it's own bespoke HLS backend to generate the RTL. The downside with Nymble is that significant effort is required to develop a high quality HLS compiler that will generate high performance RTL. Indeed, AMD have invested significantly into providing an LLVM-based HLS backend that is optimised for their architecture and so that would likely be a much more successful route long term.

Other programming models such as SYCL \cite{reinders2021data}, and implementations such as triSYCL \cite{gozillon2020trisycl} which target FPGAs, embrace a single-source, task-based approach to heterogeneous programming with FPGA support. While SYCL facilitates fine-grained control and modern C++ abstractions, its programming paradigm significantly differs from the directive-based model familiar to many scientific computing practitioners. Moreover, SYCL implementations often depend on specific hardware vendors’ toolchains, affecting portability.

Stencil-HMLS \cite{rodriguez2023stencil} provides an automated compiler flow for FPGAs based around stencils and is driven by MLIR. Lowering from ETH's \emph{stencil} dialect into a bespoke \emph{HLS} dialect, this work then integrated with that of \cite{rodriguez2023fortran} to lower the LLVM-IR into the form compatible with AMD's HLS LLVM backend. Operations in the HLS dialect are function calls, and this IR is then transformed and provided to the AMD LLVM based HLS backend. \cite{rodriguez2023stencil} demonstrated that one could gain an order of magnitude better performance than the state of the art, DaCe, without any code or algorithm changes. However this work is limited to stencils.

With the exception of the Fortran flow in \cite{rodriguez2023fortran} and Stencil-HMLS \cite{rodriguez2023stencil}, which can driven by stencils extracted from Fortran via Flang due to the work of \cite{brown2023fortran}, the work surveyed here is only compatible with C/C++. To the best of our knowledge all the existing work on integrating OpenMP with FPGAs is limited to C/C++ and there is no Fortran+OpenMP to FPGA flow.

\section{An MLIR-based OpenMP flow for FPGAs}
\label{sec:our_flow}

\begin{figure*}[htb]
\centering
 \includegraphics[width=\textwidth]{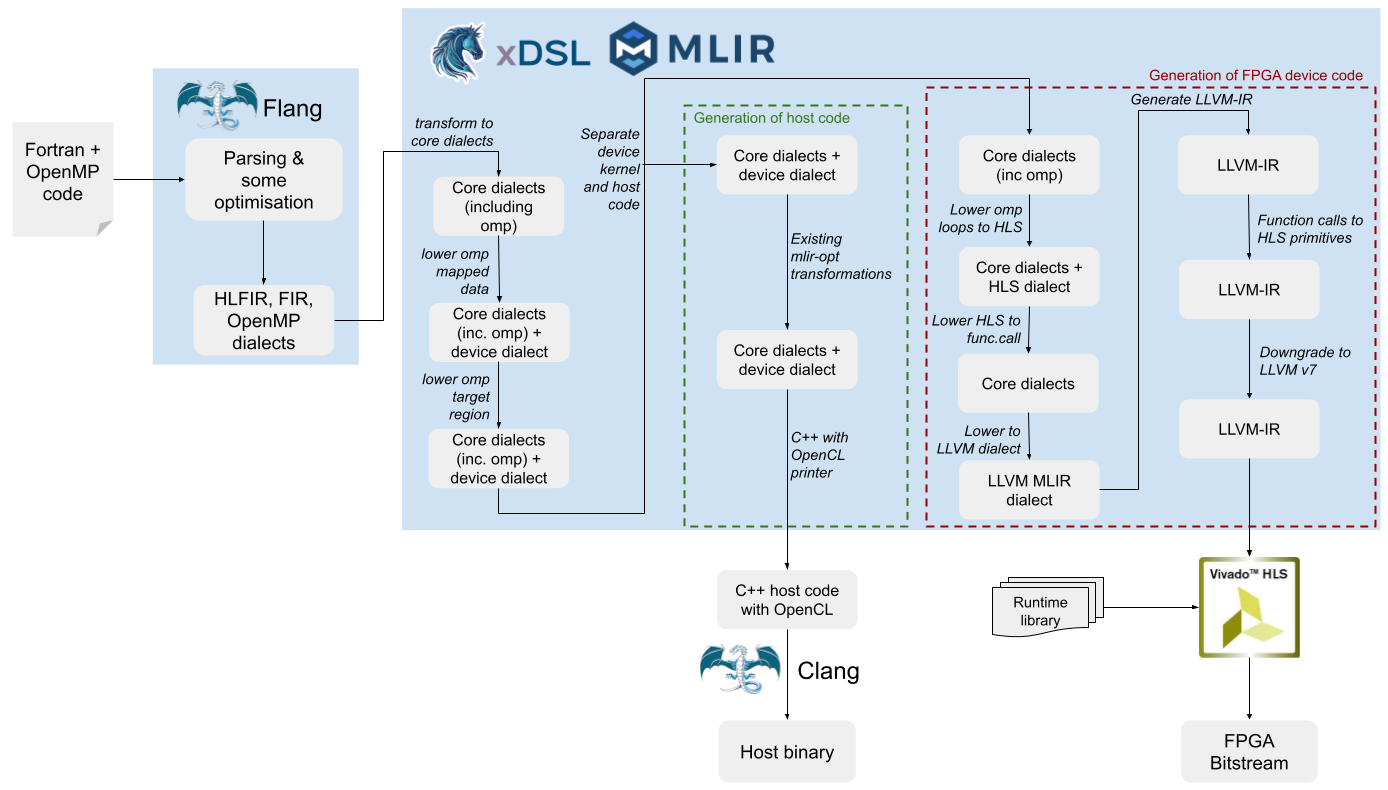}
\caption{Illustration of our compilation flow from Fortran with OpenMP to the host code and FPGA bitstream}	
\label{fig:our-flow}
\end{figure*}

Our overarching approach is to take the work of \cite{brown2024fully} which lowers Fortran, including OpenMP, to the core MLIR dialects, and then write compiler infrastructure that will connect this with the HLS MLIR dialects and transformations developed in \cite{rodriguez2023stencil}. The resulting IR will then be provided to work done in \cite{rodriguez2023fortran} to enable processing by the AMD LLVM based HLS backend. It was our hypothesis that the composability provided by MLIR would mean that these existing building blocks could be readily leveraged which would significantly reduce the complexity in developing such a compiler flow.

Figure \ref{fig:our-flow} provides a sketch of our approach where Fortran code, with the OpenMP \emph{omp} dialect, is first lowered to the core dialects using the work of \cite{brown2024fully}. We have developed a \emph{device} dialect which abstracts interaction between the host and device and ultimately then simplifies the mapping to OpenCL driver code on the host. The following list highlights the main operations in this dialect for the management of device data, and the \emph{lower omp mapped data} transformation in Figure \ref{fig:our-flow} converts from operations such as \mlirinline{omp}{map\_info} and \mlirinline{omp}{bounds\_info} into these operations.

\begin{enumerate}
    \item \mlirinline{device}{alloc} which accepts the dynamic memory sizes as operands, an identifier name and memory space as attributes. This results in a \emph{memref} allocated to a specific memory space on the device.
    \item \mlirinline{device}{lookup} which returns the memref that corresponds to the data allocated on the device based upon an identifier name and memory space.
    \item \mlirinline{device}{data\_check\_exists} returns an \emph{i1} which holds \emph{true} if the data represented by the identifier exists on the device, and \emph{false} if not.
    \item \mlirinline{device}{data\_acquire} will acquire data corresponding to a specific identifier on the device.
    \item \mlirinline{device}{data\_release} releases data corresponding to a specific identifier on the device.
\end{enumerate}

Memory on the device is tracked by a string identifier, allocated by \mlirinline{device.alloc} and then retrieved by \mlirinline{device}{lookup}. These operations return a \emph{memref} which has an associated memory space. There can be multiple memory spaces provided by a device, for instance in addition to providing 16 separate banks of HBM the U280 FPGA also contains DDR memory. The \mlirinline{memref}{dma\_start} and \mlirinline{memref}{wait} are used to transfer data between the host and device memrefs, copying data from one to the other.

What adds some complexity here is how OpenMP data is handled in the IR because there may be nested data regions. For example, the code in Listing \ref{lst:ftn_data} will first create a data region at line 4 for variable \emph{a}, with the corresponding \mlirinline{omp}{map\_info} in the IR containing the map type of \emph{from} that directs that this variable will only be copied back from the device. However, the \mlirinline{omp}{target} operation at line 5 will not only generate an \mlirinline{omp}{map\_info} operation for variable \emph{b}, with the map type of \emph{to}, but also an \mlirinline{omp}{map\_info} operation for variable \emph{a} with the map type \emph{tofrom::implicit}. This is because \emph{!\$omp target enter data} and \emph{!\$omp target exit data} which control the start and end of OpenMP data regions can be dynamic. Accordingly, OpenMP must ensure that when an offloaded loop runs on the device the data is available and \emph{tofrom} is the safest default approach.

\begin{lstlisting}[float, language=fortran, frame=lines, label=lst:ftn_data, numbers=left, caption=Example of nested OpenMP data region]
real :: a(100), b(100)
integer :: i

!$omp target data map(from:a)
!$omp target map(to:b)
    do i=1, 100
        ....
    end do
!$omp end target
!!$omp target update from(a)
!$omp end target data
\end{lstlisting}

Consequently for implicit map types the host code must be able to determine whether a device variable is already present in the data region, and if so then this implicit \mlirinline{omp}{map\_info} can be ignored. This is handled by lowering the \mlirinline{device}{data\_acquire}, \mlirinline{device}{data\_release} and \mlirinline{device}{data\_check\_exists} to operate upon an integer counter which is incremented on each \newline\mlirinline{device}{data\_acquire}, decremented on each \newline\mlirinline{device}{data\_release} and \mlirinline{device}{data\_check\_exists} checks whether the counter is greater than zero. Conditionals around \mlirinline{device}{alloc} and \mlirinline{device}{lookup}, along with \mlirinline{memref}{dma\_start} and \mlirinline{memref}{wait}, handle this for the implicit case.

The next transformation pass in Figure \ref{fig:our-flow}, \emph{lower omp target region}, transforms the \mlirinline{omp}{target} operation into operations summarised below.

\begin{enumerate}
    \item \mlirinline{device}{kernel\_create} defines a kernel and returns a handle to this.
   \item \mlirinline{device}{kernel\_launch} will launch a kernel asynchronously based upon the provided kernel handle.
   \item \mlirinline{device}{kernel\_wait} blocks until completion of a kernel.
\end{enumerate}

We undertake this transformation from \mlirinline{omp}{target} to these operations because they provide more flexibility around how kernels are scheduled and launched, as well as it being a closer representation to the OpenCL API calls that will drive FPGA execution from the host. At this point a transformation pass is executed which separates the region within the \mlirinline{device}{kernel\_create} operation into a function in a separate module. The result of this transformation pass is sketched in Listing \ref{lst:separated_mlir}, where the first module will be compiled for the host and the second for the FPGA. The data allocation and acquisition at the start of this first function can be seen, but the copying of data between the host and device via \mlirinline{memref}{dma\_start} and \mlirinline{memref}{wait} has been omitted for brevity. In this separated IR the \mlirinline{device}{kernel\_create} operation contains an empty region, as this has been extracted, with the \emph{device\_function} parameter specifying the name of the function to call on the device when the kernel is launched.

\begin{lstlisting}[style=mlir_listing, label=lst:separated_mlir, caption=Separated host and device MLIR modules.]
(*@\mlirop{builtin}{module}@*) {
   (*@\mlirop{func}{func}@*) @main() {
      %a = device.alloc(%44) <{name = "a", memory_space = 1 : i32}> : 
        (index) -> memref<100xf64, 1 : i32>
      %b = device.alloc(%44) <{name = "b", memory_space = 1 : i32}> : 
        (index) -> memref<100xf64, 1 : i32>
      device.data_acquire() <{name = "a", memory_space = 1 : i32}> : () -> ()
      device.data_acquire() <{name = "b", memory_space = 1 : i32}> : () -> ()
      ....
      %kernel = device.kernel_create(%a, %b) <{device_function = @my_kernel}> ({ 
                }) : (memref<100xf64, 1 : i32>, 
                        memref<100xf64, 1 : i32>) -> !device.kernelhandle
      device.kernel_launch(%kernel) : (!device.kernelhandle) -> ()
      device.kernel_wait(%kernel) : (!device.kernelhandle) -> ()
      ....
      device.data_release() <{name = "a", memory_space = 1 : i32}> : () -> ()
      device.data_release() <{name = "b", memory_space = 1 : i32}> : () -> ()
      func.return
   }
}

(*@\mlirop{builtin}{module}@*) attributes {target = "fpga"} {
   (*@\mlirop{func}{func}@*) @my_kernel(%a : memref<100xf64, 1 : i32>, 
                        %b : memref<100xf64, 1 : i32>) {
      ....
      func.return
   }
}
\end{lstlisting}

The second module in Listing \ref{lst:separated_mlir} contains the device IR and the attribute \emph{target} denotes which device this will be compiled for, in our case the FPGA. This module contains standard MLIR, including OpenMP IR to denote features such as parallel loops. At this point the two modules are split apart and the host code is fed into a printer that we developed which generates C++ with OpenCL that is then compiled by Clang for the host.

\begin{lstlisting}[float, language=fortran, frame=lines, label=lst:ftn_in, numbers=left, caption=Example of simple Fortran parallel loop offloaded to the FPGA]
!$omp target parallel do
    do i=1, 100
      c(i)=a(i)+b(i)
    end do
!$omp end target parallel do
\end{lstlisting}

For the device code a transformation that lowers OpenMP operations to their corresponding structure using the \emph{hls} dialect is then applied. We support OpenMP directives for fine grained control of loop parallelism, such as \textit{parallel do} and \textit{simd}. Listing \ref{lst:ftn_in} sketches a simple Fortran loop to be offloaded to the FPGA and Listing \ref{lst:ftn_mlir_hls} provides the resulting MLIR with the HLS dialect that is generated by the \emph{lower omp loops to HLS} transformation in our pipeline. The \mlirinline{hls}{interface} operation directs the mapping of kernel inputs to ports, in this case \emph{gmem0}, \emph{gmem1}, and \emph{gmem2}, and determining the protocol via passing an \mlirinline{hls}{axi\_protocol}. For this kernel each input will be mapped to a separate \emph{m\_axi} port.

\begin{lstlisting}[style=mlir_listing, label=lst:ftn_mlir_hls, caption={Sketch of generated MLIR, using the \emph{hls} dialect, from Fortran code of Listing \ref{lst:ftn_in}}]
(*@\mlirop{builtin}{module}@*) {
  (*@\mlirop{func}{func}@*) @my_kernel(%a : memref<100xf32>, %b : memref<100xf32>, 
                        %c : memref<100xf32>) { 
    %m_axi = (*@\mlirop{arith}{constant}@*) 0 : i32
    %proto = (*@\mlirop{hls}{axi\_protocol}@*)(%m_axi) : (i32) -> !hls.axi_protocol
    (*@\mlirop{hls}{interface}@*) %a, %proto {bundle="gmem0"} : (memref<100xf32>, 
                                                !hls.axi_protocol) -> ()
    (*@\mlirop{hls}{interface}@*) %b, %proto {bundle="gmem1"} : (memref<100xf32>, 
                                                !hls.axi_protocol) -> ()
    (*@\mlirop{hls}{interface}@*) %c, %proto {bundle="gmem2"} : (memref<100xf32>, 
                                                !hls.axi_protocol) -> ()
    %one = (*@\mlirop{arith}{constant}@*) 1 : i32
    %hundred = (*@\mlirop{arith}{constant}@*) 100 : i32
    %low_idx = (*@\mlirop{arith}{index\_cast}@*) %one : i32 to index
    %high_idx = (*@\mlirop{arith}{index\_cast}@*) %hundred : i32 to index
    (*@\mlirop{scf}{for}@*) %loop_idx = %low_idx to %high_idx step %low_idx {
      (*@\mlirop{hls}{pipeline}@*)(%one) : (i32) -> ()
      %array_idx = (*@\mlirop{arith}{subi}@*) %loop_idx, %one : index
      %a_val = (*@\mlirop{memref}{load}@*) %a[%array_idx] : memref<100xf32>
      %b_val = (*@\mlirop{memref}{load}@*) %b[%array_idx] : memref<100xf32>
      %result = (*@\mlirop{arith}{addf}@*) %a_val, %b_val fastmath<contract> : f32
      (*@\mlirop{memref}{store}@*) %result, %c[%array_idx] : memref<100xf32>
    }
    (*@\mlirop{func}{return}@*)
  }
}
\end{lstlisting}

After defining how kernel arguments are to be handled, several constants are then declared and converted into indexes which are used as operands to the \mlirinline{scf}{for} loop, with the \mlirinline{hls}{pipeline} operation signifying that this is a pipelined loop with the Iteration Interval (II) provided as an operand. The body of the loop will be pipelined by HLS and our transformation undertakes some simple canonicalisation to remove dependencies between loop iterations, and more complex dependency rewrite patterns can be added in the future.

In addition to the \emph{parallel do} that is illustrated in Listing \ref{lst:ftn_in}, the OpenMP \emph{simd} directive will unroll loops to run \emph{simdlen} iterations concurrently on the FPGA. Our transformation also supports the \emph{reduction} clause which separates the reduction variable into \emph{n} copies updating these copies round robin style, with the idea being that by the time this cycles around the previous update on that specific copy of the variable has been completed. These \emph{n} copies are then combined at the end depending on the reduction operation. The datatype being reduced depends on the number of copies, and this is currently determined statically by the transformation.

At this point we leverage the \emph{lower HLS to func call} transformation described in \cite{rodriguez2023stencil} to lower the \emph{hls} dialect to \mlirinline{func}{call} operations and these are then lowered by \emph{mlir-opt} to the \emph{llvm} MLIR dialect. From this dialect LLVM-IR is generated, and the work developed in \cite{rodriguez2023fortran} maps applicable function calls in the IR to corresponding AMD HLS LLVM-IR primitives, and downgrades the IR to be compatible with LLVM version 7.

The resulting LLVM-IR is provided to the AMD Vitis HLS backend along with precompiled IR containing our runtime library. The runtime library provides common functionality such as conversion between data types and reading and writing streams of data. The generated RTL is then packaged by Vitis into an IP block and run through the rest of the Vitis flow generating the FPGA bitstream. In this manner integration between the host and device code is entirely handled by the compiler, with the programmer just needing to decorate their existing Fortran loops with the appropriate OpenMP pragmas.

\section{Evaluation}
\label{sec:eval}
We chose two benchmarks from the well known scientific libraries LINPACK and LAPACK to evaluate this work. The first, 
SGESL, solves a system of linear equations of the form $Ax = b$ where $A$ has been LU decomposed and is part of the Single-precision General Matrix Factorization (SGEFA) routine. The second is SAXPY, $y = a \times x + y$, where $a$ is a scalar and $x$ and $y$ are vectors, and this represents a fundamental computation in scientific codes. We retrieved both benchmarks from the Fortran LINPACK \cite{burkardt_sgefa_openmp} and LAPACK \cite{netlib_saxpy_single_level1} repositories and added the \textit{omp target} directive to offload the compute intensive parts of the code to the FPGA and optimised the loops inside. 

Listing \ref{lst:saxpy} shows the offloaded region of the SAXPY code, where we applied a \textit{parallel do} directive to parallelise the loop with the \textit{simd} clause. We opted for vectorising the loop given all iterations are independent, where in our approach the \textit{simd} clause will unroll the loop and replicate it by the given factor. This is known as partial unrolling and, although complete unrolling could be applied to the codes when the loop bounds are known and there are enough resources available on the FPGA, partial unrolling is more common on the FPGA as it is often the \emph{sweet spot} between performance and resource utilisation. Whilst currently the unroll factor is provided by the \emph{simdlen} modifier, design space exploration could be added in the future to automatically find the best combination of directives and their parameters. Vectorisation which is applied here through loop unrolling allows for multiply-accumulate (MAC) to occur in parallel and, when the code is transformed through our flow into HLS-compatible LLVM IR, this is represented as a Vitis HLS unrolling directive. Such directives provide hints to the compiler, in this case that it should replicate the multipliers and adders involved in the computation, here by ten times.

\begin{lstlisting}[float, language=fortran, frame=lines, caption={Optimised version of saxpy in Fortran with OpenMP target offload to run on the FPGA},label=lst:saxpy]
!$omp target parallel do simd simdlen(10)
do i = 1, n
  y(i) = y(i) + a * x(j)
end do
!$omp end target parallel do simd
\end{lstlisting}

Listing \ref{lst:sgesl} shows the first offloaded loop of the SGESL subroutine. The second is optimised in a similar fashion and is omitted for brevity. This code demonstrates the simplicity of using our approach to construct a hardware pipeline from a loop in OpenMP via Fortran. The loop-carried dependency on $b$ in line 3 prevents the effective construction of a pipeline for the outermost loop, but we offload the innermost loop, via \textit{parallel do}, and as there are no dependency issues this loop will be represented effectively on the FPGA as a pipelined loop. 

\begin{lstlisting}[language=fortran, frame=lines, caption={First loop of SGESL in Fortran using OpenMP target offload to run on the FPGA}, label=lst:sgesl]
do k = 1, n - 1
  l = ipvt(k)
  t = b(l)
  if ( l /= k ) then
    b(l) = b(k)
    b(k) = t
  end if
 !$omp target parallel do
 do j=k+1,n
    b(j) = b(j) + t * a(j)
 end do
 !$omp target end parallel do
end do
\end{lstlisting}

We undertook a performance evaluation on a system containing an AMD EPYC 7502 32-Core processor with 220 GB of DRAM and an AMD Xilinx U280 FPGA. The host code was compiled with Clang 20.1.7 and bitstreams for the FPGA were built using AMD Xilinx Vitis 2020.2. The SAXPY benchmark was run with vectors of size 10K, 100K, 1M and 10M. SGESL was run with $N=256, 512, 1024, 2048$ for square matrices with dimensions $N \times N$ and vectors of size $N$. Each experiment was run a total of 10 times and the median reported here in order to account for variability. 


\begin{table*}[]
\footnotesize
\begin{tabular}{ccccl}
\cline{2-5}
\multicolumn{1}{c|}{}                  & \makecell{\textbf{N=10K} \\  \textbf{\textit{Runtime (ms)}}}       & \makecell{ \textbf{N=100K} \\  \textbf{\textit{Runtime (ms)}}}        & \makecell{\textbf{N=1M} \\  \textbf{\textit{Runtime (ms)}}}       & \multicolumn{1}{l|}{\makecell{\textbf{N=10M} \\  \textbf{\textit{Runtime (ms)}}}} \\ \hline
\multicolumn{1}{|c|}{\textbf{Fortran OpenMP}} & 1.251 $\pm$ 0.028      & 10.931 $\pm$ 0.017      & 110.245 $\pm$ 0.018     & \multicolumn{1}{c|}{1073.044 $\pm$ 0.037}       \\
\multicolumn{1}{|c|}{\textbf{Hand-written HLS}}   & 1.258 $\pm$ 0.025      & 10.925 $\pm$ 0.149      & 110.148 $\pm$ 0.018     & \multicolumn{1}{c|}{1072.888 $\pm$ 0.034}      \\ \hline
\multicolumn{1}{|c|}{\textbf{Difference}}    & +0.56\%                 & -0.06\%                   & -0.09\%                  & \multicolumn{1}{c|}{-0.01\%}        \\ \hline
\end{tabular}
\caption{Comparison of the runtime (median $\pm$ std) of our Fortran OpenMP flow against hand-written HLS code for SAXPY across different problem sizes. Runtime difference is Hand-written HLS/Fortran OpenMP.}
\label{tab:saxpy}
\end{table*}


\begin{table*}[]
\footnotesize
\begin{tabular}{ccccl}
\cline{2-5}
\multicolumn{1}{c|}{}                  & \makecell{\textbf{N=256} \\  \textbf{\textit{Runtime (ms)}}}        & \makecell{\textbf{N=512}  \\  \textbf{\textit{Runtime (ms)}}}       & \makecell{\textbf{N=1024} \\  \textbf{\textit{Runtime (ms)}}}       & \multicolumn{1}{l|}{\makecell{\textbf{N=2048}\\  \textbf{\textit{Runtime(ms)}}}} \\ \hline
\multicolumn{1}{|c|}{\textbf{Fortran OpenMP}} & 20.445 $\pm$ 0.077   & 80.791 $\pm$ 0.026   & 325.117 $\pm$ 0.116  & \multicolumn{1}{c|}{1317.247 $\pm$ 0.101}       \\
\multicolumn{1}{|c|}{\textbf{Hand-written HLS}}   & 20.594 $\pm$ 0.115   & 81.121 $\pm$ 0.023   & 325.573 $\pm$ 0.032  & \multicolumn{1}{c|}{1318.418 $\pm$ 0.042}      \\ \hline
\multicolumn{1}{|c|}{\textbf{Difference}}    & +0.73\%               & +0.41\%               & +0.14\%               & \multicolumn{1}{c|}{+0.09\%}        \\ \hline
\end{tabular}
\caption{Comparison of the runtime (median $\pm$ std) of our Fortran OpenMP flow against hand-written HLS code for SGESL across different problem sizes. Runtime difference is Hand-written HLS/Fortran OpenMP.}
\label{tab:sgefa}
\end{table*}

Tables \ref{tab:saxpy} and \ref{tab:sgefa} report the mean execution times for SAXPY and SGESL kernels, including the standard deviation, for both our Fortran OpenMP and a hand-written HLS version implemented manually in Vitis. We see that in both cases the runtime is similar, and the greatest overall difference where our Fortran OpenMP version is 6.30\% slower than the hand written HLS kernel in Table \ref{tab:saxpy} is because the standard deviation of the Fortran runtime is relatively high. The conclusion from this experiment is that the our flow does not negatively impact performance whilst significantly improving programmer productivity compared to hand-written HLS by enabling programmers to offload of Fortran loops via OpenMP to the FPGA in a seamless fashion.

\begin{table}
\centering
\begin{tabular}{|l|ccc|}
\hline
\textbf{Frontend} & \textbf{LUT \%} & \textbf{BRAM \%} & \textbf{DSP \%} \\
\hline
Fortran OpenMP    & 8.29 & 10.07 & 0.10 \\
Hand-written HLS    & 8.29 & 10.07 & 0.10 \\
\hline
\end{tabular}
\caption{Resource utilization of SAXPY in Fortran OpenMP flow compared to hand-written HLS code for problem size of N=10M.}
\label{tab:resource_util_saxpy}
\end{table}

\begin{table}[h]
\centering
\begin{tabular}{|l|ccc|}
\hline
\textbf{Frontend} & \textbf{LUT \%} & \textbf{BRAM \%} & \textbf{DSP \%} \\
\hline
Fortran OpenMP     & 8.24 & 10.07 & 0.10 \\
Hand-written HLS    & 8.22 & 10.07 & 0.23 \\
\hline
\end{tabular}
\caption{Resource utilization of SGESL in Fortran OpenMP flow compared to hand-written HLS code for problem size of N=2048.}
\label{tab:resource_util_sgefa}
\end{table}

Tables \ref{tab:resource_util_saxpy} and \ref{tab:resource_util_sgefa} report resource utilisation on the FPGA for SAXPY and SGESL respectively for the largest problem sizes. It can be seen that both flows result in the same resource utilisation for the SAXY kernel regardless. For SGESL the pattern is slightly more complex, where the amount of BRAM, the on-chip fast memory, is the same between the Fortran and Vitis versions, however LUT and DSP utilisation is different. This difference is because Vitis is able to recognise the MAC operations that are produced in the IR by the AMD HLS Clang based frontend, but it is unable to recognise similar patterns in the IR generated by our approach. This was unexpected, and clearly the HLS backend has been tuned to optimise for specific pattern that are frequently generated by their Clang frontend but less effort has been undertaken by AMD engineers to generalise this. Consequently, the HLS backend maps the MAC operations to DSP slices in the hand-written HLS version, but to LUTs in the Fortran version. However, the it can be seen from Table \ref{tab:sgefa} that the use of DSPs by the Vitis version does not provide any performance advantage here. Improving the IR generated to fit the MAC pattern expected by Vitis and memory optimisations will be addressed by future work.

Tables \ref{tab:saxpy_power_cpu} and \ref{tab:sgefa_power} report median power draw on the FPGA comparing our approach offloading Fortran OpenMP and hand-written HLS also against a single core of the CPU. It can be seen that both approaches on the FPGA result in similar power draw numbers, which is around half that of a single CPU core. Consequently it can be observed that our approach maintains the low power benefits of FPGAs compared with running on the CPU.

\begin{table}[]
\footnotesize
\begin{tabular}{ccccl}
\cline{2-5}
\multicolumn{1}{c|}{}                  & \makecell{\textbf{N=10K} \\  \textbf{\textit{Power (W)}}}        & \makecell{\textbf{N=100K}  \\  \textbf{\textit{Power (W)}}}       & \makecell{\textbf{N=1M} \\  \textbf{\textit{Power (W)}}}   & \multicolumn{1}{l|}{\makecell{\textbf{N=10M} \\  \textbf{\textit{Power (W)}}}} \\ \hline
\multicolumn{1}{|c|}{\textbf{Fortran OpenMP}} & 21.847     & 23.528      & 25.535      & \multicolumn{1}{c|}{24.167} \\ 
\multicolumn{1}{|c|}{\textbf{Handwritten HLS}}   & 22.178      & 22.496    & 23.998 & \multicolumn{1}{c|}{24.297} \\ 
\multicolumn{1}{|c|}{\textbf{CPU single core}}     & 56.13    & 55.08    & 57.31     & \multicolumn{1}{c|}{54.91} \\ \hline
\end{tabular}
\caption{Median power draw of our Fortran OpenMP flow compared to hand-written HLS code for SAXPY across different problem sizes, with CPU results included.}
\label{tab:saxpy_power_cpu}
\end{table}


\begin{table}[]
\footnotesize
\begin{tabular}{ccccl}
\cline{2-5}
\multicolumn{1}{c|}{}                  & \makecell{\textbf{N=256} \\  \textbf{\textit{Power (W)}}}        & \makecell{\textbf{N=512}  \\  \textbf{\textit{Power (W)}}}       & \makecell{\textbf{N=1024} \\  \textbf{\textit{Power (W)}}}   & \multicolumn{1}{l|}{\makecell{\textbf{N=2048} \\  \textbf{\textit{Power (W)}}}} \\ \hline
\multicolumn{1}{|c|}{\textbf{Fortran OpenMP}} & 21.866      & 22.989      & 24.243     & \multicolumn{1}{c|}{24.278} \\ 
\multicolumn{1}{|c|}{\textbf{Handwritten HLS}}   & 22.363           & 23.121    & 23.640     & \multicolumn{1}{c|}{24.066} \\ 
\multicolumn{1}{|c|}{\textbf{CPU single core}}     & 52.70         & 53.71     & 52.44         & \multicolumn{1}{c|}{52.82 } \\ \hline
\end{tabular}
\caption{Median power of our Fortran OpenMP flow compared to hand-written HLS code for SGESL across different matrix sizes.}
\label{tab:sgefa_power}
\end{table}

Table \ref{tab:loc} reports the lines of code across the dialects and transformations that we leverage in this work and described in Section \ref{sec:our_flow}. It can be seen that the work reported here, to connect the core dialects with the HLS dialect is 2363 lines of code and even the total number of lines across all these components is very modest compared to what would be required by bespoke compilers sitting outside of the MLIR ecosystem.

\begin{table}[h]
\centering
\begin{tabular}{|c|c|}
\hline
\textbf{Component} & \textbf{Lines of code}\\
\hline
OpenMP to HLS dialect (this work) & 2363 \\
HLS dialect and lowering from \cite{rodriguez2023stencil} & 2382 \\
Integrating LLVM and AMD HLS backend \cite{rodriguez2023fortran} & 1654  \\
Lowering from HLFIR \& FIR to core dialects \cite{brown2024fully} & 5956 \\
\hline
\end{tabular}
\caption{Lines of code in MLIR dialects and transformations that are leveraged in this work}
\label{tab:loc}
\end{table}

\section{Conclusions and further work}
\label{sec:conclusions}
In this paper we have explored the development of an MLIR-based compiler flow for offloading Fortran OpenMP loops to FPGAs. Whilst OpenMP offload to FPGAs has been studied in detail before, to the best of our knowledge this work is the first to integrate this with MLIR and also the first to provide such a capability from Fortran. This paper acts as a case study to demonstrate the benefits of MLIR and how the flexibility and composability of the ecosystem significantly reduces the effort in developing complicated compiler flows. Indeed, we believe that this provides additional evidence to the HPC community to argue that it should continue investing in MLIR and building tools that leverage and enrich the ecosystem.

Our compilation flow respects OpenMP data semantics and translates directives into HLS-specific pragmas compatible with Vitis HLS. Leveraging, in the main, core MLIR dialects our approach supports any MLIR-compatible front end, and by situating FPGA acceleration inside the MLIR ecosystem our methodology provides a unified and extensible framework that bridges the gap between existing HLS tools and directive-based programming models. Unfortunately the Polygeist C MLIR frontend does not support OpenMP target offload, but as the CIR project which is integrating Clang with MLIR becomes more mature it should be trivial to integrate that work with our approach here. Fundamentally, MLIR enables us to reduce dependency on proprietary compilers and promotes portability across hardware and front ends, representing a significant step towards mainstream adoption of FPGAs in HPC environments.

The work presented in this paper matures support in MLIR for targeting FPGAs which can be extended in the future. Whilst we have focused in this paper on driving from Fortran and a subset of OpenMP pragmas, the MLIR dialect and transformations that we have developed, along with the existing building blocks being leveraged, are generic and can be extended to support a wider range of the OpenMP standard in future. We will also look to integrate with OpenACC which also has a corresponding MLIR dialect, this should be made significantly easier given the work undertaken in this paper and previously.

\begin{acks}
This work has been funded by an EPSRC Doctoral Training Partnership (EP/T517884/1) and supported by a Royal Society of Edinburgh personal research fellowship award number 3271. The CPU runs in this paper ran on NextGenIO which was funded by the EU Horizon 2020 research and innovation programme under grant agreement No 671591. For the purposes of open access, the author has applied a CC BY public copyright license to any Author Accepted Manuscript version arising from this submission.
\end{acks}

%
%
%
\bibliographystyle{ACM-Reference-Format}
\bibliography{references.bib}

\end{document}